%% file: main.tex
\documentclass[a4paper,twoside]{article}

\input{preamble/preamble}

\input{preamble/macros}

\begin{document}

\title{
\vspace{-1.8cm}
\begin{center}
\small
\textit{Published in the Proceedings of the 23rd International Conference on Security and Cryptography - Volume 1: SECRYPT, SciTePress, pages 959-970. DOI: \url{https://doi.org/10.5220/0015000000004103}}
\end{center}
\vspace{0.5cm}
Birds of a Feather Flock Together: Network-Based Detection of Coordinated Disinformation Campaigns on Telegram}

\author{\authorname{Panteleimon Tsagkarakis\sup{1,2}, Emmanouil Papadogiannakis\sup{1,2}, and Evangelos Markatos\sup{1,2}}
\affiliation{\sup{1}Foundation for Research and Technology - Hellas (FORTH), Heraklion, Greece}
\affiliation{\sup{2}University of Crete, Heraklion, Greece}
}


\input{sections/0.abstract}

\onecolumn \maketitle \normalsize \setcounter{footnote}{0} \vfill

\input{sections/1.introduction}
\input{sections/2.background}
\input{sections/3.methodology}
\input{sections/4.findings}
\input{sections/5.related_work}
\input{sections/6.conclusion}

\bibliographystyle{apalike}
\bibliography{main}

\appendix
\input{sections/appendix}

\end{document}

%% file: preamble/preamble.tex
\usepackage{epsfig}
\usepackage{subcaption}
\usepackage{calc}
\usepackage{amssymb}
\usepackage{amstext}
\usepackage{amsmath}
\usepackage{amsthm}
\usepackage{multicol}
\usepackage{pslatex}
\usepackage{apalike}
\usepackage{algorithm2e}
\usepackage[bottom]{footmisc}
\usepackage[hidelinks]{hyperref}
\usepackage{booktabs}
\usepackage{framed}
\usepackage{enumitem}
\usepackage{xcolor}
\usepackage{SCITEPRESS}

%% file: preamble/macros.tex
\newcommand{\eg}{e.g.,\xspace}

\newcommand{\ie}{i.e.,\xspace}

\newcommand{\point}[1]{\par\smallskip\noindent\textbf{#1}}

\newcounter{finding}
\renewcommand{\thefinding}{\arabic{finding}}

\newcommand{\finding}[1]{
  \refstepcounter{finding}
  \begin{leftbar}
  \noindent\textbf{Finding~\thefinding:}~#1  
  \end{leftbar}
}

%% file: sections/0.abstract.tex
\abstract{
In recent years, disinformation has increasingly proliferated across social networks.
The firing of fact-checkers from Meta and the disbanding of Twitter's Trust and Safety Council suggest that this trend will continue to escalate.
While disinformation sources (\eg social network accounts) can sometimes be identified, new accounts emerge daily, making tracking a moving target.
In this work, we propose a graph-based methodology to discover previously unknown Telegram accounts that spread disinformation.
Starting from a set of verified disinformation groups, we examine their interconnections and identify new accounts that contribute to the dissemination of false narratives.
Our approach is language-agnostic, as it relies solely on structural relationships between accounts rather than analyzing their message content.
Using this approach, we identify 37 previously unknown disinformation channels (a threefold increase).
We demonstrate that Telegram channels display extremely dogmatic behavior with up to 80\% of messages being labeled as propaganda.
Our findings reveal that misinformation on Telegram spreads within tightly interconnected clusters, in some cases, with over 86K identical messages being shared to multiple channels, suggesting coordinated disinformation campaigns.
}

%% file: sections/1.introduction.tex
\section{\uppercase{Introduction}}
\label{sec:introduction}

Disinformation has emerged as a central challenge in the digital information ecosystem, shaping public opinion across political, social, and health issues.
The majority of American adults already regard fabricated stories as confusing or difficult to discern~\cite{fakeNewsConfusion}.
In fact, during the 2016 and 2020 U.S. elections, the dissemination of false or misleading claims online was pervasive, raising concerns about its potential influence on the electoral outcome~\cite{lakhta,teensInBalkans}.
To make matters worse, in 2020, misinformation related to COVID-19 is estimated to have resulted in nearly 6,000 hospitalizations worldwide~\cite{whoStudy}, and at least 800 deaths~\cite{islam2020covid}, underlying the fact that disinformation endangers the health and lives of people exposed to it.

Recent legislative efforts have sought to curb online disinformation, yet their effectiveness remains limited.
Governments and regulators have introduced laws to increase platform accountability, enhance content moderation transparency, and penalize the deliberate dissemination of false information~\cite{lim2023chilling}.
However, these measures struggle to keep pace with the speed and scale at which misinformation circulates across digital networks~\cite{papadogiannakis2025before}.
The problem has been further exacerbated by social media platforms, through actions such as Meta's termination of fact-checking teams~\cite{facebookFactCheckers} and Twitter's disbanding of its Trust and Safety Council~\cite{twitterTrustSafety}.
As a result, despite numerous legal initiatives, misinformation continues to thrive, underscoring the gap between regulatory frameworks and their practical effect.

Unlike traditional media, social media platforms enable virtually anyone to produce and spread false information~\cite{grinberg2019fake,kansaon2025fake}, with two-thirds of social media experts fearing that such platforms amplify disinformation~\cite{socialMediaExperts}.
In recent years, Telegram has become a prominent communication platform, offering encrypted messaging, large channels, and relative anonymity, features that have also facilitated the spread of disinformation~\cite{telegramFakeNews}.
Shortly after the escalation of the war in Ukraine, a fake Telegram account impersonating president Zelenskyy urged Ukrainian soldiers to surrender, appearing even on television before being debunked~\cite{ukraineTelegramNPR}.
In Poland, over 22 Polish‑language Telegram channels amplified pro‑Russian narratives, including fabricated stories about Ukrainian refugees, influencing public opinion and deepening societal divides~\cite{polandTelegram}. 
Yet the internal structure of Telegram propaganda ecosystems, and how new channels emerge and gain traction, remain under explored. 
Manual moderation cannot keep pace as new accounts appear daily, with over 1 billion monthly active users in 2025~\cite{telegramDilemma}.

In this work, we introduce a language-agnostic, graph-based method that uncovers previously unknown Telegram disinformation spreaders utilizing only their inter-channel references. 
Starting from 18 known disinformation channels, we apply our method to real-world data and uncover twice as many additional disinformation spreaders.
Our proposed method overcomes language barriers that require analysts to understand the shared content or even the idioms of the language.
Unlike text, which is easily manipulated, graph-based representations offer a comprehensive, domain-independent view that reveals underlying relationships and coordinated behaviors.

\begin{figure*}[t]
  \centering
  \includegraphics[width=0.8\textwidth]{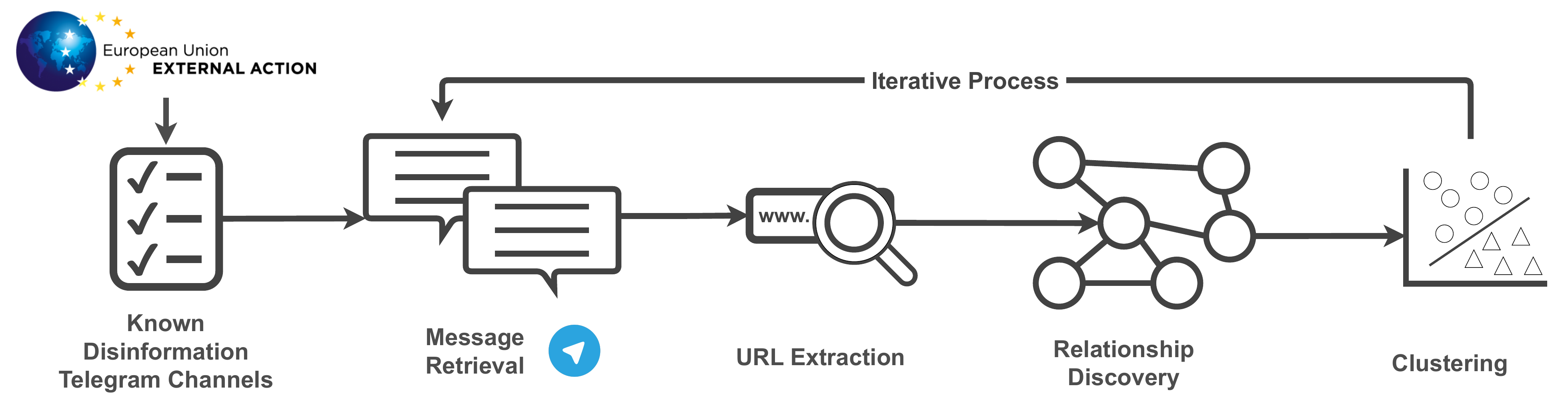}
  \caption{Overview of the overall graph-based methodology to detect disinformation channels on Telegram.}
  \label{fig:methodology}
\end{figure*}

Overall, the contributions of this work are:
\begin{enumerate}
    \item We propose a language-agnostic and graph-based methodology that discovers disinformation channels on Telegram focusing only on their sharing behavior. We demonstrate the efficacy of our methodology with 37 newly discovered channels being labeled as propaganda spreaders. We establish that our language-agnostic approach uncovers patterns of coordinated behavior among accounts spreading false narratives.
    \item Our analysis reveals that disinformation on Telegram circulates within echo chambers, where the narratives of a small number of channels are disproportionately amplified and forwarded by others. Certain channels flood over 3,000 distinct messages that are subsequently reshared word-for-word across other channels. We find over 2,000 distinct messages to be identical across more than 4 distinct disinformation channels, indicating coordinated dissemination campaigns.
    \item We identify channels that leverage platforms like YouTube and Twitter to expand their reach, with over half (52.5\%) of their content removed for policy violations and more than 84\% referencing known fake news websites.
    \item We uncover temporal patterns in Telegram campaigns, identifying more than 7K instances of different channels sharing near-simultaneous identical content. Coordinated bursts in posted messages, spanning over 71 clusters of channels, tend to align with major political events, highlighting strategic timing and coordinated influence efforts.
    \item We provide public access to the Telegram channels we identify as disinformation spreaders~\cite{openSource}.
\end{enumerate}

%% file: sections/2.background.tex
\section{\uppercase{Background}}
\label{sec:background}

Telegram is a cloud‑based messaging platform which blends three different communication modes: 
(i) private chats with optional end-to-end encryption, 
(ii) groups hosting up to hundreds of thousands of participants, and 
(iii) channels broadcasting messages to an unlimited audience.
Channels display posts in reverse‑chronological order and include a one‑tap ``Forward'' button that copies messages and attachments to a new destination, while also preserving the original author.
These features make sharing content more convenient, but they also make it easier for large amounts of unverified information to spread quickly.

These design choices matter because dis-/mis-information thrives in environments where content publication is effortless, sharing is encouraged, and users are able to form ideologically homogeneous communities.
Modern platforms, including Telegram, satisfy all these conditions, and its privacy-centric branding may empower those spreading misinformation by limiting accountability.

Within Telegram, two channel types are especially relevant for studying information quality.
First, \emph{group‑style channels} act like large public forums, where users can freely post messages, allowing reactions to accumulate quickly and discussions to spiral.
By contrast, \emph{broadcast‑style channels} resemble one‑way blogs or news outlets.
Specifically, only the owner or operator is allowed to publish messages, while subscribers consume the content passively.
Since each post originates from a single voice, narratives remain consistent and can be repeated without challenge, fostering coordinated disinformation.

In this work, we focus on \emph{broadcast‑style channels} to study disinformation on Telegram.
This one-to-many structure resembles the behavior of traditional newspapers or news websites, enabling the dissemination of information without debate.
Unlike group chats or group-style channels, where many users contribute, broadcast channels have specific operators, allowing us to trace the exact origin of false narratives.

%% file: sections/3.methodology.tex
\section{\uppercase{Methodology}}
\label{sec:methodology}

We set out to discover Telegram channels that spread disinformation and false narratives in a coordinated manner.
Disinformation campaigns on Telegram are difficult to combat due to the platform's manual moderation and limited user data~\cite{kireev2025characterizing}.
To detect disinformation channels and differentiate them from other legitimate information sharing channels, we propose a language-agnostic, graph-based detection methodology that relies on online resources shared across channels.
We provide an overview of our methodology in Figure~\ref{fig:methodology}.
Inspired by previous work (\eg~\cite{willaert2022disinformation,kireev2025characterizing,la2025tgdataset}), our methodology starts from a set of known disinformation channels, fetches the messages shared in these channels and iteratively discovers new channels that are frequently promoted.
Finally, we cluster detected channels spreading false information to uncover patterns of coordinated dissemination.
In contrast to conventional fact-checking approaches, our language-agnostic methodology eliminates the need for text translation.
Rather than examining linguistic traits, we represent information as relationships within graphs, focusing on the underlying connections among Telegram channels.

\subsection{Detection \& Data Collection}

Our methodology discovers new Telegram disinformation channels by identifying channels frequently promoted or referenced by known disinformation spreaders.
The initial set of disinformation spreaders is extracted from the analysis of the European External Action Service (EEAS) on Foreign Information Manipulation and Interference Threats~\cite{EEAS}.
This analysis presents threat actors involved in over 100 incidents of information manipulation, including 18 Telegram channels.
All these channels are attributed to Russia, even if they use different languages (\eg English), or if they don't target Russian domestic residents (\eg Russian embassy in Japan).
We use this list of channels as a seed to our methodology.

Using the official Telegram API, we fetch messages posted in these Telegram channels, including message text, media metadata, author handles and outbound links.
We then analyze all messages to extract outbound URLs, using them to identify which other channels a given channel promotes or forwards.
Moreover, we implement some data cleaning mechanisms to ensure that the references we capture are valid.
All shortened URLs, including \texttt{bit.ly}, \texttt{goo.gl}, and \texttt{t.me/} codes, are expanded.
We also remove self-references and references to private groups, and ignore any URLs that fail to resolve.

Using all extracted URLs, we follow a graph-based methodology to represent relationships and information sharing between Telegram channels.
Each node in our graph represents a distinct Telegram channel.
When one channel references or forwards a message from a different channel, we introduce a directed edge connecting the two nodes.
The more messages are shared across the two channels, the greater the edge weight.
For instance, if the Telegram channel \texttt{X} regularly forwards and mentions messages from channel \texttt{Y}, an edge with increased weight will connect the two nodes.
Edge weights are normalized by dividing each channel's reference count by the maximum reference count.
That is, the weight of the edge from \texttt{X} to \texttt{Y} is calculated as the number of times \texttt{X} referenced \texttt{Y}, divided by the highest number of references \texttt{X} made to any single channel.
By construction, edge weights range from 0 to 1, with a maximum value of 1.
This representation measures the relative strength of interaction between channels and illustrates how much attention one channel gives to the other.

Next, we perform edge pruning to retain only the edges with the highest weights and establish the most important relationships among channels.
For each channel, we only retain the top 5\% connections and drop any edges that result from fewer than 10 messages.
This threshold is selected based on prior work that applied the same value using a similar methodology~\cite{papadogiannakis2022leveraging}.
This choice is further motivated by minimizing false positives, that is, channels that are frequently mentioned but not semantically associated.
Following prior work~\cite{luceri2024unmasking}, we perform node pruning, dropping any isolated nodes and keep only the top 20\% of nodes based on their in-degree.
This step measures not just whether a channel is mentioned, but how significant those mentions are compared to others, signaling information alignment or even coordinated information dissemination.
We empirically evaluate multiple threshold values to examine their effect on the graph construction.
We select the aforementioned thresholds as they provide the most stable and consistent behavior.
Higher threshold values result in increasingly sparse graphs, which substantially increase the computational cost.
Once channels closely linked to the initial disinformation group are identified, we iteratively repeat message collection and relationship discovery (as illustrated in Figure~\ref{fig:methodology}).
Each iteration adds new nodes to the graph until there are no new channels to be integrated.
Finally, to discover coordinated disinformation campaigns, we group Telegram channels into clusters based on mutual references, using the Louvain community detection algorithm.

\begin{figure}[t]
  \centering
  \includegraphics[width=\columnwidth]{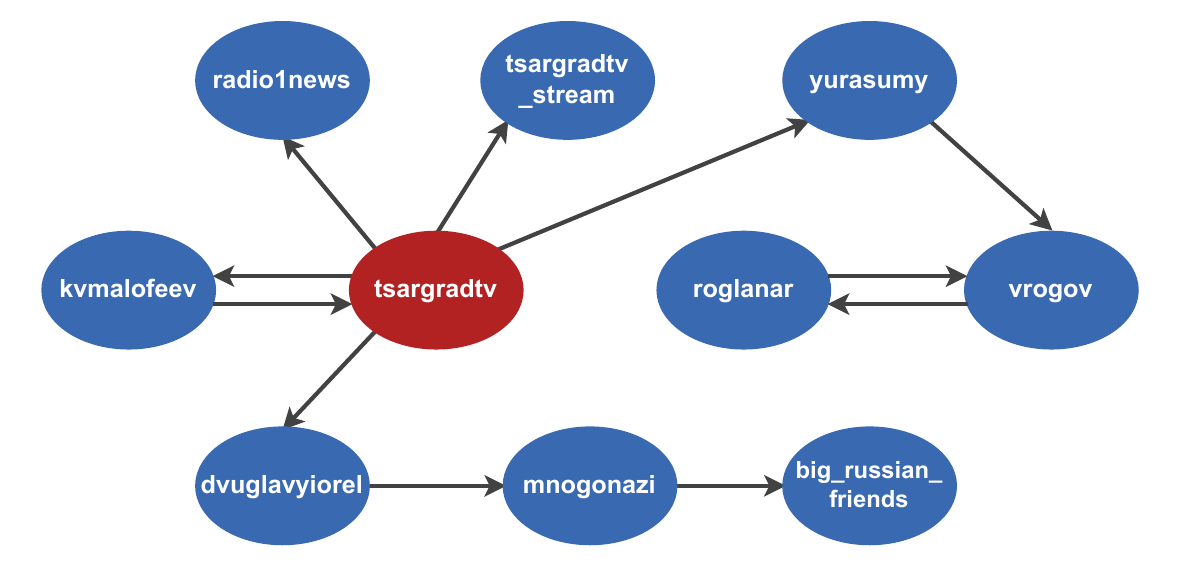}
  \caption{Network of Telegram channels clustered according to how frequently they reference each other. Distinct clusters indicate groups of channels that communicate or cite one another more often, suggesting communities of information exchange. The channel initially identified by the EEAS as a disinformation source is highlighted in red.}
  \label{fig:exampleCluster}
\end{figure}

Starting from the seed list of 18 Telegram channels associated with disinformation, we fetch and process all Telegram messages posted in channels up until 2024-12-31.
We discover over 1,800 distinct cross channel references (\ie forwarding of other posts) and our methodology identifies 55 disinformation Telegram channels grouped into 9 clusters of information alignment.
We provide a cluster example in Figure~\ref{fig:exampleCluster} and make our list of identified disinformation channels publicly available to foster further research~\cite{openSource}.

\subsection{Disinformation Classification}

Starting from a list of known disinformation spreaders, our methodology identifies previously unrecognized channels that exhibit strong informational alignment.
The validation in this section is performed on all 55 channels in our final dataset: the 18 EEAS seed channels and the 37 newly discovered channels.
These channels frequently reference each other to promote similar narratives or cite one another to reinforce claims.
Such patterns of content overlap suggest that the newly detected channels also disseminate mis-/dis-information.
To verify this hypothesis, we follow a two-prong approach using separate evaluation frameworks.
First, we make use of Propasafe~\cite{sharma2025propasafe}, an LLM-based framework, trained on real news articles, that detects propaganda and labels sentences as ``\texttt{Neutral}'', ``\texttt{Mild Propaganda}'' and ``\texttt{Severe Propaganda}''.
We deploy Propasafe locally and feed it with messages from the detected Telegram channels.
Each message is split into sentences while quotes longer than two words are zero weighted, so only original speech is evaluated.
Altogether, we feed $\sim$2M messages to Propasafe and discover that a staggering 24.08\% of them are labeled ``\texttt{Mild Propaganda}''.
To make matters worse, there are over 88K messages (4.57\%) considered ``\texttt{Severe Propaganda}''.
We discover over 550K propaganda messages across all Telegram channels our methodology identifies as disinformation spreaders.

In addition to this, we deploy OpenAI's ChatGPT-4o model to evaluate the trustworthiness of our methodology and produce a structured diagnostic report for each Telegram channel.
We instruct the model to form a report based on source reliability, fact-checking indicators, emotional or manipulative language, political or ideological bias and use of anonymity.
The full prompt can be found in Appendix~\ref{sec:chatGptPrompt}.
We feed a random sample of messages (200-300 messages for each group, depending on the message sizes) from each Telegram channel to the model and discover that all but two groups are classified as ``Propaganda'' or ``Fake News''.
The other two channels are deemed as ``Misinformation-Prone'', while none of them achieved a trustworthy rating.

To ensure the reliability of the model's outputs, two authors of this work independently evaluated the analytical reports generated for each channel.
The two authors were allowed to use any translation tool and assessed whether the model's classification and rationale are consistent with observable patterns in the selected messages.
Their independent assessments resulted in full agreement with the model's decisions, suggesting that the ChatGPT-4o identification process and the Propasafe classification align with human judgment.
Additionally, as a control, to verify the correctness of models, we also fed messages from the Telegram channels of mainstream and reliable news outlets (\eg The New York Times, Bloomberg), education groups (\eg \texttt{BooksMania}), video games (\eg \texttt{Catmoonity}) and hobbies (\eg \texttt{supercarscontent}).
Models correctly identify such channels as reliable sources, indicating that they can successfully distinguish professional journalism from disinformation.
Although LLMs are not definitive ground-truth classifiers, agreement between two independent models, along with human verification, strongly supports that our method effectively uncovers Telegram channels spreading disinformation.

%% file: sections/4.findings.tex
\section{\uppercase{Interconnected Ecosystems}}
\label{sec:findings}

Our content-agnostic, graph-based methodology reveals the underlying topology of disinformation propagation on Telegram.
This section examines relationships between channels, focusing on content sharing, disinformation spreading, and temporal coordination.

\subsection{False Claims \& Propaganda}

Our detection methodology identified 55 Telegram channels as disinformation spreaders.
We further investigate these channels and investigate their false claims.
Using Propasafe~\cite{sharma2025propasafe}, we study the volume of propaganda in Telegram channels.
We find that channels directly disseminate propaganda messages to large audiences.
To our surprise, we discover that 12 channels have more propaganda messages than neutral ones.
In Table~\ref{tab:propagandaGroups}, we highlight the 5 Telegram channels that published the largest number of messages labeled as propaganda, along with the number of subscribers (as of November 2025).
In all cases, the portion of propagandistic content exceeds that of neutral messages, indicating that these channels are consistently engaged in disseminating propaganda rather than neutral information.
The channel \texttt{kvmalofeev} demonstrates the most extreme behavior with 79.67\% of overall messages spreading either mild or severe propaganda.
With each channel reaching thousands to over a million subscribers, these active channels demonstrate a strong focus on misleading messaging, reflecting their role in sustaining specific false narratives.
Finally, we find that, as of November 2025, channels \texttt{mnogonazi} and \texttt{tsargradtv} are no longer accessible from the authors' country due to local legal restrictions.

Moreover, we randomly sample some messages with manipulative content and manually analyze their content.
We utilize online tools to translate messages to English and discover that Telegram channels use techniques commonly used in fake news to manipulate readers~\cite{chen2021persuasion}.
First, we identify messages that selectively quote specific sections of public speeches, ignoring the broader context and demonstrating political bias.
This behavior was increasingly evident in messages of \texttt{SputnikGeorgia} against the EU and USA.
Generally, we uncover strong alignment with anti-Western ideologies in the channels we study.
Moreover, we find that language is often emotionally charged and divisive while it often celebrates violence (\eg ``July's harvest of victorious missile strikes'' published by \texttt{SputnikLive}).
Quotation marks are often used to discredit opposing opinions (\eg ``recognized'', ``peaceful transfer'').
Blanket accusations and unsubstantiated claims are common, with references often relying on untrustworthy or affiliated sources (\ie borrowed credibility).

\finding{
Telegram channels serve as vectors for misleading and manipulative content, where emotionally charged language and sensationalized narratives amplify unverified claims.
}

\begin{table}[t]
\caption{Telegram channels with the highest proportions of propagandistic content. These channels exhibit the greatest ratios of messages labeled as mild or severe propaganda. Over half of the published messages contain propagandistic elements, indicating concentrated disinformation activity.}
\centering
\scriptsize
\begin{tabular}{lrr}
\toprule
{\textbf{Telegram Channel}} & {\textbf{\% of Propaganda}} & {\textbf{\# of subscribers}}\\
& \textbf{Messages} \\
\midrule
     \texttt{kvmalofeev}          & 79.67\% & 1.26M \\
     \texttt{EvPanina}            & 68.01\% & 222K \\
     \texttt{kremlebezBashennik}  & 66.42\% & 65K \\
     \texttt{RossiyaNeEvropa}     & 59.37\% & 12K \\
     \texttt{mnogonazi}           & 58.28\% & 358K \\
\bottomrule
\end{tabular}
\label{tab:propagandaGroups}
\end{table}

\begin{figure*}[t]
    \begin{minipage}[t]{0.32\textwidth}
        \centering
        \includegraphics[width=\columnwidth]{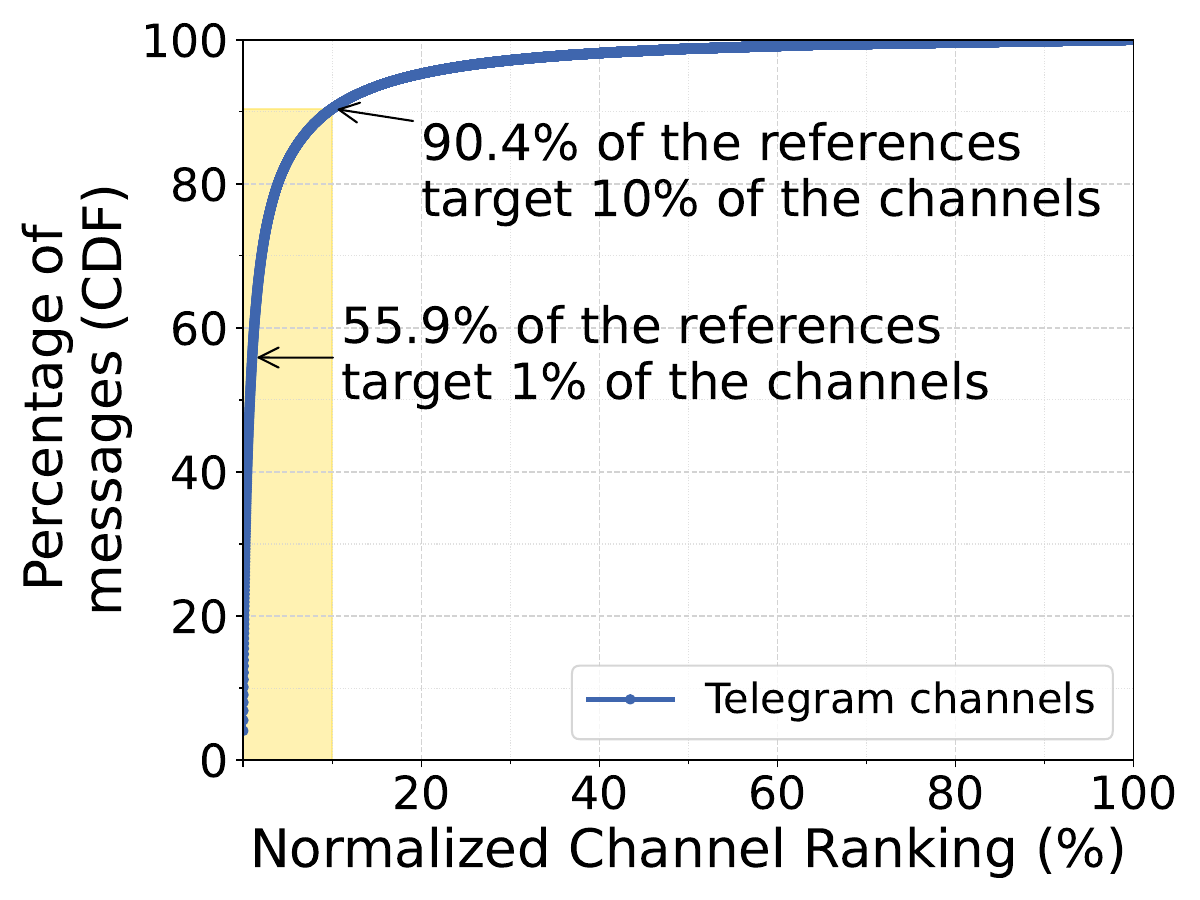}
        \caption{Distribution of sharing frequency among Telegram channels. Channels are ranked in descending order of received references. The resulting distribution follows the Pareto principle, with a small number of channels (10\%) dominating the high frequency region (90\% of shared messages).}
        \label{fig:messageSharePowerlaw}
    \end{minipage}
    \hfill
    \begin{minipage}[t]{0.32\textwidth}
        \centering
        \includegraphics[width=\columnwidth]{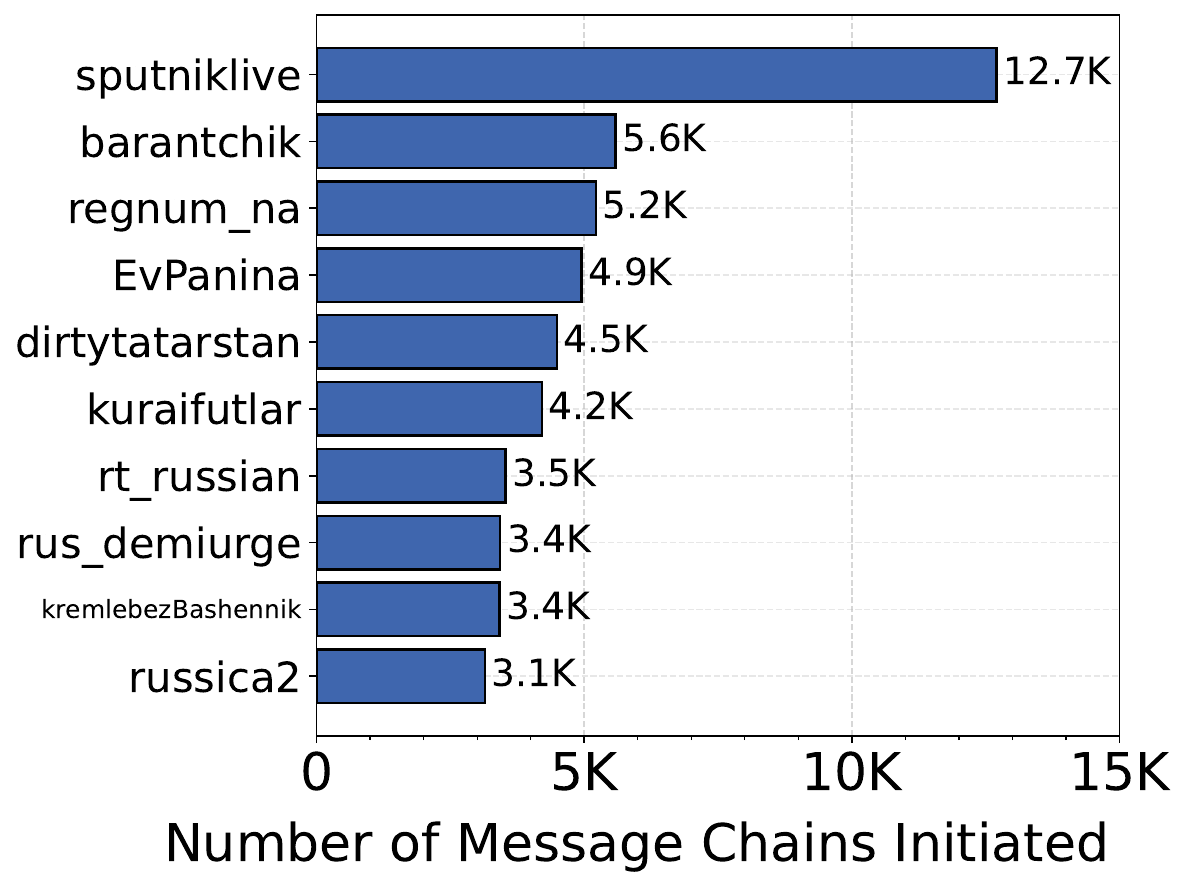}
        \caption{Distribution of identical message chains initiated by disinformation channels. A small number of channels are responsible for initiating a disproportionately large share of identical message chains, suggesting that duplication behavior is concentrated within specific communication sources.}
        \label{fig:topMessageChains}
    \end{minipage}
    \hfill
    \begin{minipage}[t]{0.32\textwidth}
        \centering
        \includegraphics[width=\columnwidth]{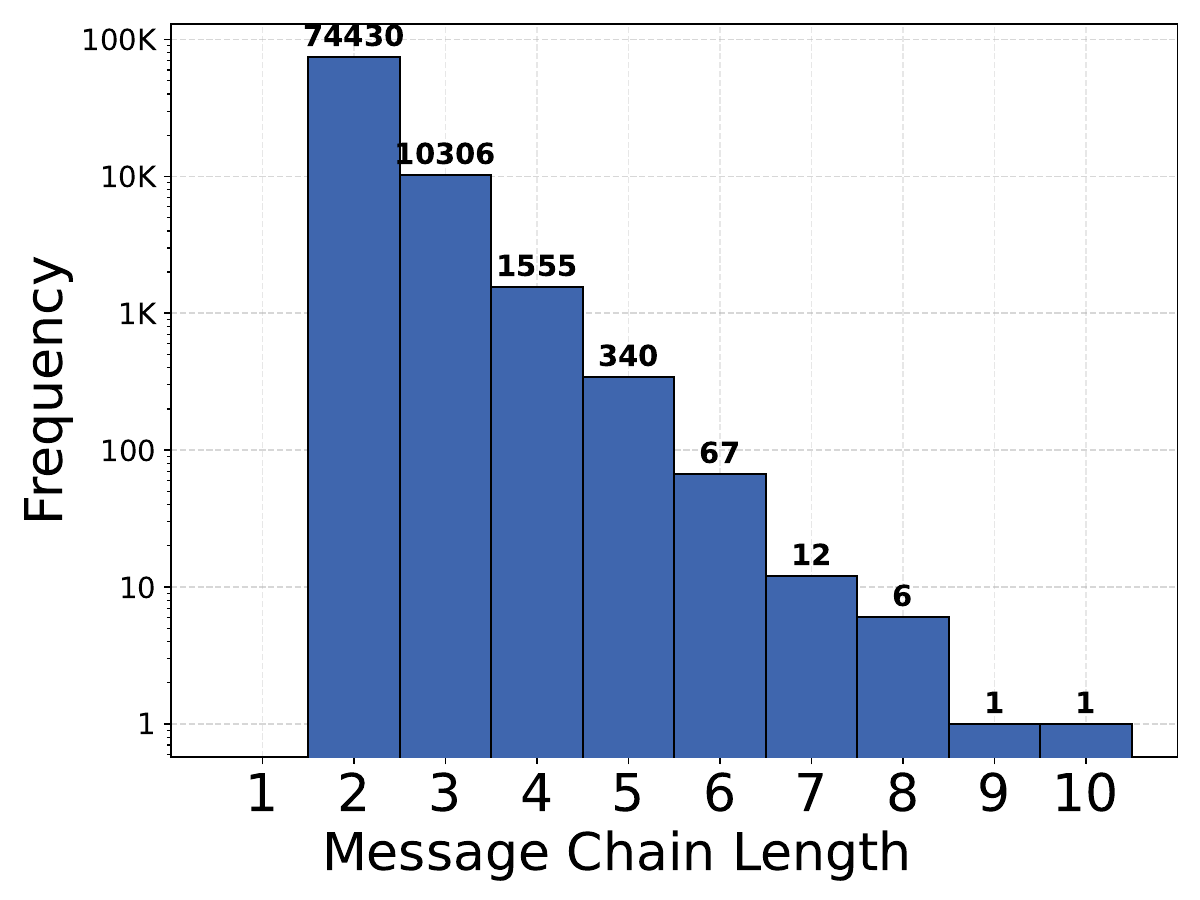}
        \caption{Distribution of length of identical message chains ($y$-axis in log scale). The distribution is heavily skewed, with the vast majority of chains being short, while several chains exhibit considerably greater lengths. While identical messages are typically limited in scope, they can still spread widely.}
        \label{fig:messageChainLength}
    \end{minipage}
\end{figure*}

\subsection{The Echo Chamber Effect}
\label{sec:echochambers}

In Telegram, the channel functionality along with limited moderation, often foster tightly coupled communities where opposing views are systematically excluded. 
This behavior results in \emph{echo chambers}: environments in which users are predominantly exposed to information that reinforces their preexisting beliefs.
To test this hypothesis, we analyze the patterns of cross-references among Telegram channels, evaluating whether users are exposed to isolated narratives.

First, we study messages that the discovered 55 disinformation channels share and repost.
For each disinformation channel, we count references to other channels, 
regardless of whether those channels are themselves labeled as disinformation.
We then aggregate the references each channel receives and rank channels in descending order of reference popularity.
Thus, the highest ranked channel represents the most frequently referenced source among disinformation channels.
We plot in Figure~\ref{fig:messageSharePowerlaw} the distribution of message references in Telegram channels.
We observe that message sharing follows the Pareto Principle.
That is, 90\% of the messages that disinformation channels share come from just the top 10\% of channels, implying a highly skewed distribution of influence among the channels.
Additionally, $\sim$56\% of forwarded messages come from just 1\% of the most popular channels (based on references).
A small subset of channels is driving the vast majority of content circulation and the top channels play a disproportionate role in shaping narratives.
Operators of the detected disinformation channels share and forward hundreds or even thousands of messages from specific groups, while sharing very few messages from the majority of other channels.
These results indicate that message sharing is not random but rather follows a structured approach.
Operators of these channels focus their dissemination efforts on specific source channels, repeatedly amplifying content from specific groups while minimally interacting with other channels.
Such behavior underscores a structured and selective strategy in content propagation.
This finding is inline with prior work that has demonstrated that 5\% of the channels are responsible for 40\% of the forwarded messages in fringe communities on Telegram~\cite{hoseini2024characterizing}.
The observed pattern, where a channel disproportionately amplifies a small, ideologically aligned subset of sources suggests the presence of echo chambers.

Figure~\ref{fig:messageSharePowerlaw} demonstrates that there is a structural isolation between channels.
However, to demonstrate that these channels are ideologically aligned and that the content they share is homogeneous, we examine the actual posted messages.
We focus on messages that propagate verbatim between channels, leading to repeated narratives.
To discover such messages, we treat every post's plaintext body as a hashable string and detect duplicates across the entire corpus.
Each message is normalized by removing newlines, converting all text to lowercase, and concatenating paragraphs or text segments.
We also exclude messages of five words or fewer to remove content lacking significant meaning.
We define a ``message chain'' as a sequence of identical or near-identical messages that appear across multiple Telegram channels.
We discover over 159K messages that are reshared word-for-word across at least two disinformation channels, forming 86K distinct message chains and suggesting coordinated disinformation campaigns.
Messages are broadcast across multiple channels to amplify reach and reinforce credibility through repetition.

For every message belonging to a message chain, we identify the channel that first published the text according to timestamp order.
This allows us to rank groups based on how many message chains they initiate.
We plot in Figure~\ref{fig:topMessageChains} the most prolific message chain originators, all of which start at least 3K distinct message chains.
These 10 channels are responsible for 59\% of the message chains we discover.
We observe that the channel \texttt{sputniklive} is the dominant channel, initiating over 12K distinct message chains, accounting for \texttt{13.74}\% of all chains in our dataset.
It is evident that a handful of sources supply the majority of recyclable content, reinforcing the idea of echo chambers, where few distinct voices dominate.
Since each chain represents a specific message reposted verbatim across multiple channels, initiating a chain corresponds to producing content that will subsequently propagate widely across Telegram.
The dominance of the channels in Figure~\ref{fig:topMessageChains} indicates that they play a key role in driving the spread of information within their network, functioning as major sources whose content is repeatedly amplified by other channels.

We measure the length of message chains, defined as the number of distinct Telegram channels that repost the same message verbatim.
We plot in Figure~\ref{fig:messageChainLength} the distribution of message chain lengths revealing that nearly 2,000 distinct messages are reposted in at least four different channels.
This pattern indicates broad repetition across information networks, which may reflect coordinated dissemination of specific content and highlights messages that achieve wide visibility across multiple channels.
Studying the time window of message duplications we find that the median message is reposted within 1 hour and 7 minutes 
while about 8.5\% of messages are shared within just 1 minute, suggesting either automated dissemination or coordinated messaging.

\finding{
Disinformation Telegram channels exhibit bias by disproportionately sharing messages from a small set of other channels and repeating identical messages across multiple channels, thus leading to the formation of echo chambers.
}

\subsection{Misinformation Spreaders}

Next, we study the URLs of external domains shared among the Telegram channels we discover.
We cross-reference these external links with a list of misinformation websites from an academic publication~\cite{papadogiannakis2023funds}.
We find that $\sim$84\% of the Telegram channels we discover have shared at least one article from the list of misinformation websites.
Moreover, we identify three Telegram channels that drive the majority of misinformation references.
Specifically, the channel \texttt{rus\_demiurge} has shared articles from almost 120 distinct misinformation websites.
Channels \texttt{EvPanina} and \texttt{warfakes} have shared multiple articles from $\sim$70 and $\sim$30 distinct misinformation websites, respectively.
These channels act as hubs for disinformation spreading and expose their users to false narratives.

In addition to this, we observe that disinformation channels rarely remain confined to Telegram, using other online platforms to expand their reach and credibility.
Telegram channels often share YouTube videos related to political events, global health issues or even war clips.
We collect all URLs redirecting users to YouTube and collect the IDs of channels that posted the videos.
We identify over 1,200 distinct YouTube URLs and detect that more than half of them (52.5\%), resolve to channels that have since been banned from YouTube for policy violations.
Among the resolvable channels, the most frequent are nationalist commentary streams \texttt{ssmysly} and \texttt{metametrica\_live}. 
The increased number of banned channels suggests attempts to spread controversial or prohibited material, including misinformation or misleading content\footnote{\url{https://support.google.com/youtube/answer/9288567}}.

Finally, we identify that disinformation channels often use platforms like Odysee, VK, Facebook and X (formerly Twitter) to support their claims or to amplify specific narratives.
We find that \texttt{sputniksrbija}, a disinformation channel with over 59K members is the one that shares the most messages from X (formerly Twitter), for broadcasting and quick circulation of short messages.
Additionally, we find that disinformation channels often use the video hosting platform Odysee to promote false narratives.
Odysee is a decentralized video platform that has been found to host content for hate groups and extremists because of its lack of moderation~\cite{odysee}.
Our analysis shows that the channel \texttt{sputniksrbija} has shared more than 53K Odysee videos over a period of 28 months.
This ecosystem of amplification makes moderation efforts extremely difficult, since removing a claim on one platform does not prevent its persistence on others.

\finding{
Disinformation channels on Telegram utilize other social media platforms like YouTube, X (formerly Twitter), Odysee along with fake news websites to amplify their narratives, maximize their reach and increase their perceived credibility. The content they share is often banned from other platforms due to policy violations.
}

\begin{figure}[t]
  \centering
  \includegraphics[width=0.9\columnwidth]{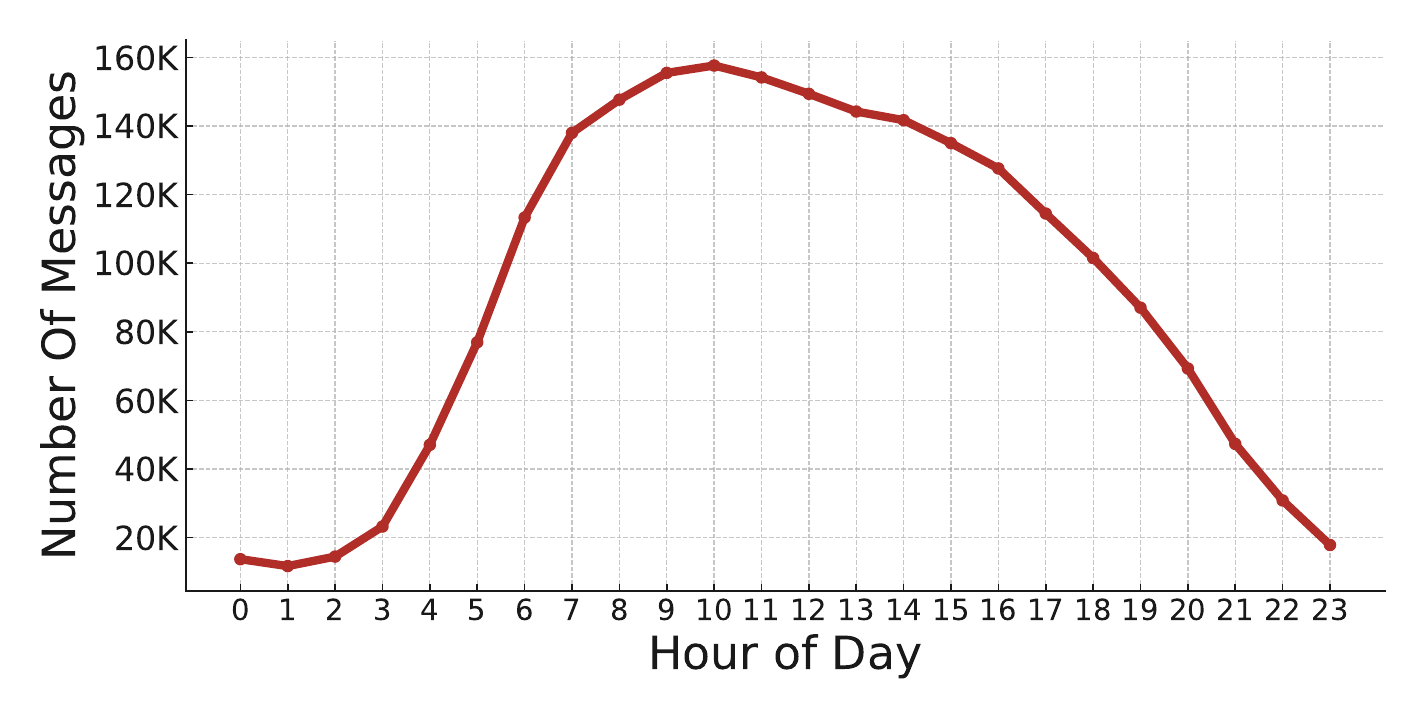}
  \caption{Distribution of message postings on Telegram disinformation channels. The majority of messages ($\sim$57\%) are posted between 07:00 and 17:00, indicating that disinformation activity follows a regular daytime schedule rather than occurring uniformly throughout the day.}
  \label{fig:messageHours}
\end{figure}

\subsection{Coordinated Activity Patterns}

To better understand the dissemination strategies of known Telegram disinformation channels, we analyze their posting patterns.
First, we study the weekdays that Telegram disinformation channels post messages.
We find that the posting volume is greater on weekdays compared to weekends, with an average 34\% increase in the number of messages posted on weekdays.
We study and plot in Figure~\ref{fig:messageHours} the distribution of posting times for messages on Telegram disinformation channels.
All message timestamps are converted to Moscow Standard Time (MSK), as all identified channels originate from Russia.
We discover that posting has a regular pattern with most messages being posted from 07:00 to 17:00, accounting for 56.87\% of all messages sent.
This regularity is important because it suggests that disinformation activity is follows predictable daily schedules, rather than occurring randomly across the day.
It may reflect targeting of working-hour audiences to maximize reach or the operational routines of channel operators.

\begin{figure}[t]
  \centering
  \includegraphics[width=\columnwidth]{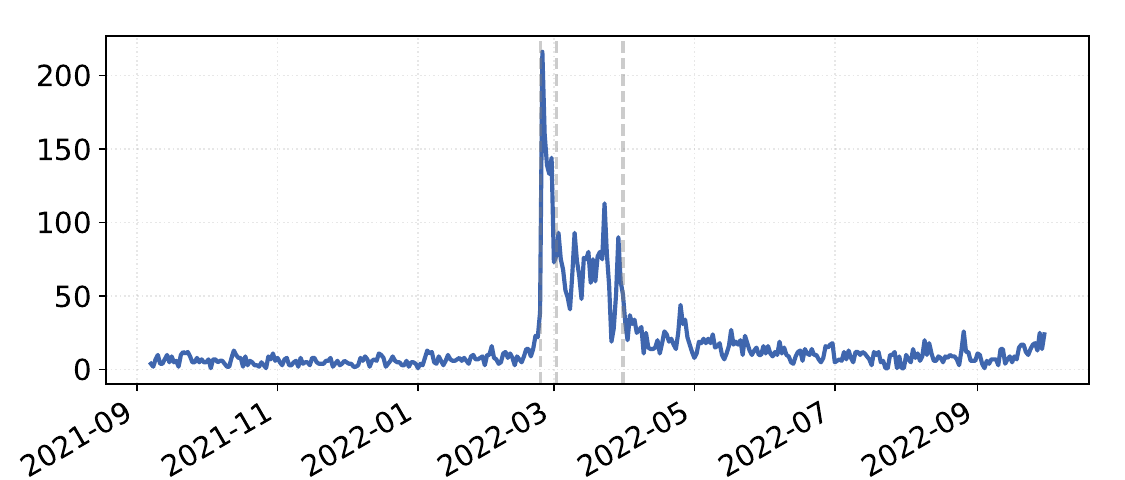}
  \caption{Change point analysis of daily posting activity in disinformation channel \texttt{barantchik}. The analysis identifies distinct shifts in posting frequency over time, with pronounced spikes corresponding to major political events. These irregularities suggest that the channel's activity intensity is event-driven rather than stable across periods.}
  \label{fig:peltAnalysis}
\end{figure}

Next, we study individual number of posts per day for each channel, exploring peaks in posting activity that may coincide with political events or breaking news.
To detect correlation between real world events and the posting patterns of Telegram disinformation channels, we perform a change point analysis utilizing the pruned exact linear time (PELT) method~\cite{killick2012optimal}.
We apply the PELT method to all discovered channels and further examine their change point dates.
We discover that disinformation campaigns often intensify around key events, with the number of messages posted in Telegram channels greatly increasing.
In Figure~\ref{fig:peltAnalysis}, we plot the change point analysis on the disinformation channel \texttt{barantchik}, 
illustrating how activity responds to key external events.
We observe that major political or news events correspond to significant spikes in daily posting volume.
The greatest spike is observed on February 24, 2022, the day that Russia invaded Ukraine, starting a war, with a $4.5\times$ increase in the number of messages that day.
Similar behavior is also observed in other channels (\eg \texttt{kuraifutlar} and \texttt{neoficialniybezsonov}) while the Crocus City Hall attack~\cite{crocusCityHall} also created major spikes in posting activity.
Altogether, the change point analysis provides evidence of increased activity synchronized with political events, suggesting that the posting activity is event-driven rather than constant.

\begin{figure}[t]
    \centering
    \includegraphics[width=0.9\columnwidth]{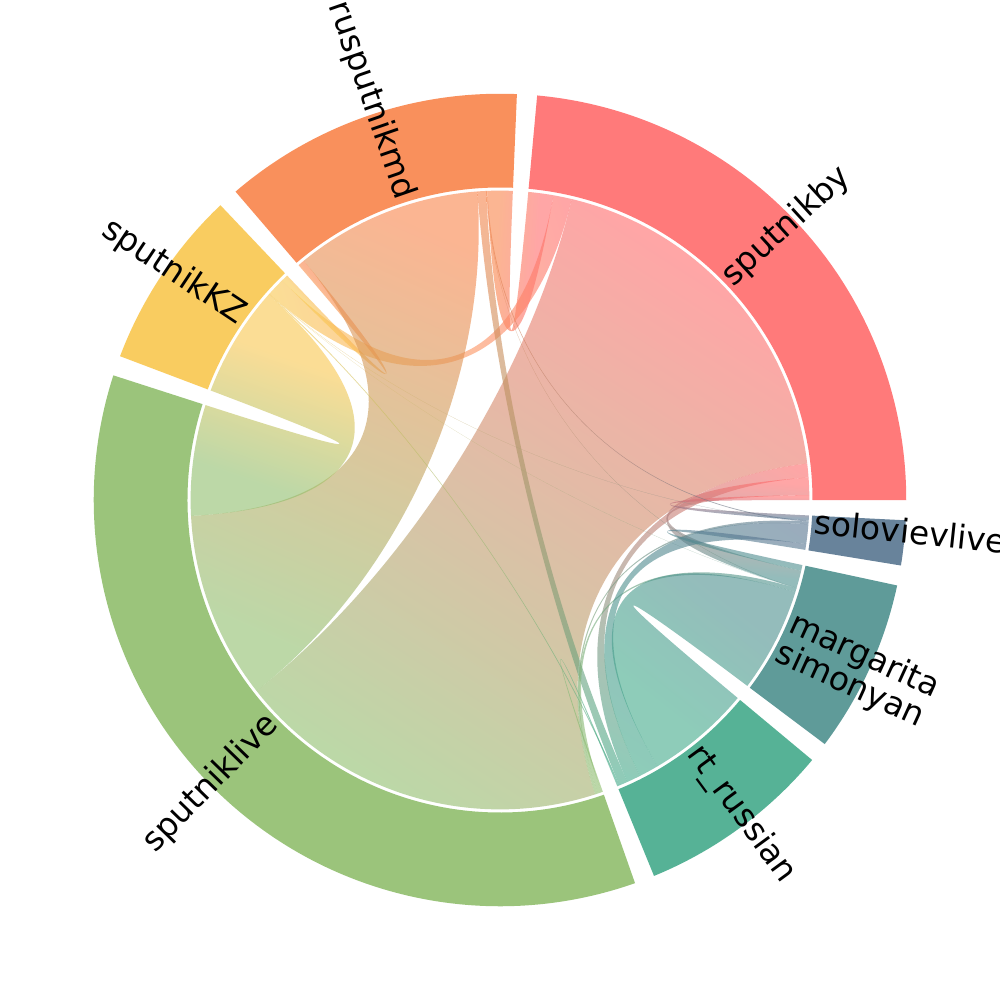}
    \caption{Cross channel synchronization of identical message postings on Telegram. Several channels exhibit a consistent pattern of concurrent posting, suggesting coordinated dissemination strategies. The pairing between \texttt{sputniklive} and \texttt{sputnikby} accounts for more than 3K identical messages shared within a one-minute interval.}
    \label{fig:identicalMessages}
\end{figure}

Finally, we take a closer look at the actual time that each message is sent and focus on chains of identical messages (as defined in Section~\ref{sec:echochambers}).
We discover deliberate posting bursts coordinated across multiple channels.
By comparing posting times, we find over 7.7K instances of messages being reposted across at least two different channels within one minute.
For instance, on April 9, 2023, an extensive message talking about the birthday of the Executive Secretary of the Commonwealth of Independent States, was posted verbatim on the channels \texttt{sputnikKZ}, \texttt{sputnikby}, \texttt{sputniklive} and \texttt{rusputnikmd} within just 20 seconds.
Similarly, on May 31, 2023, the seemingly unrelated channels \texttt{DvuglavyiOrel}, \texttt{kvmalofeev} and \texttt{tsargradtv} posted exactly the same message within just 7 seconds.
The channel of Margarita Simonyan, a Russian media executive, regularly posts almost simultaneously identical messages with the TV network RT, Sputnik Belarus and the TV channel Soloviev.Live.
We plot in Figure~\ref{fig:identicalMessages}, the identical messages shared by different pairs of disinformation channels that have been sent within one-minute period.
We observe that specific channels have a common strategy of posting the same content concurrently, with the greatest pair being that of \texttt{sputniklive} and \texttt{sputnikby} that have posted over 3,300 identical messages within just one minute.
Altogether, we find 71 distinct clusters of Telegram channels that have posted at least 10 identical messages within one minute.
These findings indicate that there is centralized and automated control among channels.
They also highlight orchestrated amplification, allowing messages to rapidly appear across multiple channels and increase their visibility.

\finding{
Disinformation channels exhibit temporal and behavioral coordination patterns, with thousands of identical messages often appearing across multiple channels within seconds, and overall posting activity showing spikes of up to $4.5\times$ more messages in correlation with major events.
}

%% file: sections/5.related_work.tex
\section{\uppercase{Related Work}}
\label{sec:related}

Coordinated disinformation campaigns~\cite{magelinski2020detecting}, disinformation for profit~\cite{papadogiannakis2023funds}, social botnets~\cite{fazil2020socialbots}, and Foreign Information Manipulation and Interference (FIMI)~\cite{prysiazhniuk2025strategic,pante2025beyond} have become a recognized security threat, prompting government bodies such as the European External Action Service (EEAS) to catalog tactics, actors, and recurring narratives across platforms~\cite{EEAS}.
In the context of the 2024 U.S. presidential election, groups of accounts worked together deceptively and across multiple social media platforms like X, Facebook, and Telegram~\cite{cinus2025exposing}.
Coordinated Inauthentic Activity (CoIA) research shows cross-platform coordination, especially between X and Telegram, promoting Russian-affiliated media, low-credibility websites, and conspiratorial content.
Using large language models and graph analysis to trace 2024 U.S. election discussions on Telegram,~\cite{paoletti2025tracing} found highly polarized conversations organized into distinct communities.
In fact, fake news websites often exhibit synchronized uptime periods and serve identical content during these times, particularly around significant events like the 2016 U.S. presidential election~\cite{chalkiadakis2021rise}.

In~\cite{pacheco2021uncovering}, authors studied coordinated social media campaigns, showing that individual accounts may appear normal in isolation.
Using a similar methodology to ours, state-sponsored information operations on Twitter were studied in~\cite{luceri2024unmasking}.
By building similarity networks from user behaviors and pruning nodes, this method detected malicious accounts with over 95\% precision and can predict future campaign involvement.
Building on this idea,~\cite{papadopoulos2023fndaas} proposed an automated system to detect fake news websites without analyzing article content, focusing instead on structural and network features like domain age, DNS records, IP links, and third-party ad requests.

The rise of instant messaging has drawn research attention on technical and social aspects, with Telegram standing out for large group channels and strong encryption.
Telegram's network is fundamentally different from other social networks as it is extremely sparse and fragmented into many disconnected components~\cite{dargahi2017analysis}.
Additionally, a large portion of COVID-19 misinformation on Telegram was shared through images and videos, not just text~\cite{sosa2022multimodal}.
In~\cite{la2023sa}, authors studied Telegram fake channels, which impersonate public figures, companies, and services to deceive users, spread conspiracy theories, or run scams.
Contrary to our work, using a manually labeled dataset of known official and fake channels, they trained a machine learning model to automatically detect fake channels.

In~\cite{kireev2025telegram}, authors highlighted the challenges of studying propaganda on Telegram, where malicious messages are often deleted by moderators and thus missed by researchers.
TGDataset, the largest publicly available collection of Telegram channels, comprising over 120,000 channels and 400 million messages, was created to provide a comprehensive snapshot of the platform's ecosystem~\cite{la2025tgdataset}.
Similar to our methodology, the authors started with a diverse set of seed channels and expanded it by following forwarded messages to discover new ones.
A newer dataset introduced in~\cite{blas2025unearthing}, contained over 43 thousand chats and over 1 billion messages focusing on the 2024 US Presidential Election.
A similar approach was followed in~\cite{willaert2022disinformation}, where authors studied the network of conspiracy Dutch-speaking Telegram channels.
This methodology allowed them to discover communities with overlapping narratives.
Along the same lines as our work,~\cite{kireev2025characterizing} uncovered two coordinated networks spreading pro-Russian and pro-Ukrainian propaganda.
The authors proposed a novel detection method that analyzes the relationship between a user's original message and the propaganda account's reply.
Their approach resists evasion by relying on legitimate-user content and remains effective as propaganda topics change.
Finally, previous work demonstrated that even though false claims are more commonly shared on Telegram, credible news sources attract more views~\cite{herasimenka2023misinformation}.

%% file: sections/6.conclusion.tex
\section{\uppercase{Discussion \& Conclusion}}

\point{Summary:}
The rise of social media platforms has significantly altered the dynamics of information dissemination, with Telegram emerging as a particularly influential platform.
Bad actors utilize Telegram to spread false narratives, conspiracy theories and coordinated manipulative content.
In this work, we study the disinformation ecosystem on Telegram and propose a graph-based and language-agnostic detection methodology that discovers disinformation clusters based on their interdependence and message dissemination. 
We find that Telegram channels disproportionately share messages from other politically aligned channels and use emotional language and sensationalized narratives to spread false narratives.
Such channels serve as echo chambers, where misleading information is reinforced and rapidly circulated.
Additionally, we uncover that disinformation spreaders regularly use other social media platforms and known disinformation websites to amplify their reach and acquire false credibility.
Finally, through a timeline analysis of posted messages, we demonstrate that there is coordinated activity and strategic efforts to amplify specific narratives across different Telegram channels, particularly during major political events.

\point{Discussion:}
Telegram's emphasis on privacy limits the opportunities for content moderation, allowing false or misleading content to circulate unchecked.
Disinformation on Telegram circulates in echo chambers, reinforcing false narratives.
The high replication of identical messages across multiple channels suggests coordinated dissemination strategies, possibly automated or orchestrated.
Such activity is usually implemented with synchronized posting across channels, often within seconds.
These techniques create the illusion of widespread agreement, increasing both the visibility, as well as the perceived credibility of misleading information. 
Furthermore, the fact that Telegram channels leverage YouTube or other platforms (\eg Twitter and Odysee) underscores the interconnection of the disinformation ecosystem, making mitigation even harder if all efforts remain on a single platform.
This highlights the need for cross-platform coordination among social media companies, regulators, and fact-checking organizations.
\point{Limitations \& Future Work:}
While our methodology uncovers previously unidentified disinformation channels, it captures only a subset of Telegram's disinformation ecosystem, as private groups and ephemeral content remain inaccessible.
Despite insights gained in this work, further research is required to fully understand and counter disinformation.

\point{Ethical Considerations:}
To ensure transparency and reproducibility, we report specific Telegram channels analyzed in this study.
All channels referenced are publicly accessible, and some were identified in a public report by the European External Action Service.
No personally identifiable information is disclosed.
Mentioning these channels is necessary to illustrate observed patterns of disinformation and to support the validity of our findings, rather than to single out individual users or promote their content.

\section*{\uppercase{Acknowledgments}}
Funded by the European Union (project ATHENA 101132686).
Views and opinions expressed are however those of the authors only and do not necessarily reflect those of the European Union or the granting authority.
Neither the European Union nor the granting authority can be held responsible for them.

%% file: sections/appendix.tex
\section{ChatGPT Evaluation Prompt}
\label{sec:chatGptPrompt}

\noindent
\textbf{Criteria for Analysis (Text-Only):}
\begin{itemize}[wide, labelwidth=!, labelindent=0pt,nosep]
    \item Source Reliability: Are the claims backed by credible sources? Are references provided, and are they legitimate?
    \item Fact-Checking Indicators: Do the messages contain verifiable or debunked false information? Compare against established fact-checking sources if necessary.
    \item Emotional \& Manipulative Language: Are the messages written in a way that provokes strong emotions (fear, anger, outrage) rather than providing neutral information?
    \item Repetition \& Echo-Chamber Effects: Are the same claims repeated frequently without new supporting evidence?
    \item Logical Fallacies \& Misinformation Tactics: Are there any common techniques used to spread misinformation, such as straw‑man arguments, false equivalences, cherry‑picking data, or misleading statistics?
    \item Political or Ideological Bias: Is there a strong alignment with a political agenda, and does it present one-sided narratives without acknowledging opposing views?
    \item Call to Action \& Mobilization: Do the messages attempt to mobilize people toward specific actions (\eg protests, boycotts) based on unverified claims?
    \item Use of Anonymity \& Lack of Accountability: Are the sources of the messages anonymous, or do they discourage verification from external sources?
\end{itemize}

\noindent
Overall Verdict: Based on the above, categorize the group into one of the following:
(i) Reliable Information: No signs of fake news or propaganda, (ii)  Misinformation-Prone: Some questionable claims but not necessarily intentional propaganda, or (iii) Propaganda/Fake News: Clear manipulation, misinformation, or deceptive intent.